\documentclass[showpacs,preprint,amsmath,amssymb,nofootinbib]{revtex4-1}

\usepackage{amsmath}
\usepackage{graphicx}
\usepackage{epsf}
\usepackage{psfrag}
\usepackage{epsfig}
\usepackage{graphics}
\usepackage{amsfonts}
\usepackage{epstopdf}

\newcommand{\be}{\begin{equation}}
\newcommand{\ee}{\end{equation}}
\newcommand{\bq}{\begin{eqnarray}}
\newcommand{\eq}{\end{eqnarray}}

\begin{document}

\title{(Un)determined finite regularization dependent quantum corrections: the Higgs decay into two photons and the two photon scattering examples}

\date{\today}

\author{A. L. Cherchiglia$^{(a)}$} \email[]{adriano@fisica.ufmg.br}
\author{L. A. Cabral$^{(b)}$} \email[]{cabral@uft.edu.br}
\author{M. C. Nemes$^{(a)}$}\email[]{mcnemes@fisica.ufmg.br}
\author{Marcos Sampaio$^{(a)}$} \email []{msampaio@fisica.ufmg.br}

\affiliation{(a) Departamento de F\'isica - ICEx - Universidade Federal de Minas Gerais\\ P.O. BOX 702, 30.161-970, Belo Horizonte - MG - Brazil}
\affiliation{(b) Departamento de F\'isica - Universidade Federal do Tocantins\\ P.O. BOX 132, 77804-970, Araguaina - TO - Brazil}
\begin{abstract}

\noindent

We investigate the appearance of arbitrary, regularization dependent parameters introduced by divergent integrals in two \textit{a priori} finite but  superficially divergent amplitudes: the Higgs decay into two photons and the two photon scattering. We use a general parametrization of ultraviolet divergences which makes explicit such ambiguities. Thus we separate in a consistent way using Implicit Regularization the divergent, finite and regularization dependent parts of the amplitudes which in turn are written as surface terms. We find that, although finite, these amplitudes are ambiguous before the imposition of physical conditions namely momentum routing invariance in the loops of Feynman diagrams.  In the examples we study momentum routing invariance turns out to be equivalent to gauge invariance. We also discuss the results obtained by different regularizations and show how they can be reproduced within our framework  allowing for a clear view on the origin of regularization ambiguities.

\end{abstract}


\maketitle

\section{Introduction}
\label{intro}

On the last 4th July a new boson was announced using its decay into two photon as one of the main channels of discovery \cite{:2012gk,:2012gu}. The immediate question that arose was whether this new boson corresponds to the one predicted by the Standard Model (Higgs boson) or not. To help answering this question theoretical predictions (loop corrections) on such decays must be set on consistent grounds.

Some time ago the W loop calculation of the Higgs decay into two photons was performed in the unitary gauge and the result obtained \cite{Gastmans:2011wh} contradicted previous ones found in the literature \cite{Ellis:1975ap,Ioffe:1976sd,Shifman:1979eb}. The reason pointed by the authors was the use of Dimensional Regularization (DReg). Soon after many authors performed calculations in the framework of Dimensional Regularization \cite{Marciano:2011gm}, Lattice \cite{Bursa:2011aa} and Loop Regularization \cite{Huang:2011yf}. In all cases the old results were recovered shedding many doubts on the statements presented in \cite{Gastmans:2011wh}. Other authors used Cut-off Regularization \cite{Piccinini:2011az,Shao:2011wx} obtaining the same result of \cite{Gastmans:2011wh} thus concluding that such regularization is non-predictive if one works on the unitary gauge. Other works were devoted to the discussion of the decoupling theorem \cite{Shifman:2011ri,Jegerlehner:2011jm} questioning the reliability of the predictions made in \cite{Gastmans:2011wh}. 

Contemporary to Gastmans et al. work, another paper questioned an old-established result in the literature: the cross section of the two photon scattering \cite{Kanda:2011vu}. Once again, doubts were raised against the use of regularization. A work followed in which this issue was explained \cite{Liang:2011sj} in the framework of Dimensional and Pauli-Villars Regularization recovering the old results found in the literature \cite{Karplus:1950zz,Karplus:1950zza}.

The aim of the present work is to revise these two calculations with the purpose of illustrating that
\textit{a priori} undefined quantum corrections in Feynman
diagram calculations, which  entail regularization scheme dependence, are the common denominator of such discussion.  Such arbitrarinesses must not be mistaken by finite  parameters related to the freedom of defining renormalization constants to be fixed by renormalization conditions
(i.e. the choice of a renormalization point). We propose a general
parametrization valid at arbitrary
loop order to handle such ambiguities which acts on the physical dimension of the theory thus being particularly useful to dimensional specific models. Moreover an alternative exhibition of such arbitrariness in terms of arbitrary n-loop integrals is proposed. In this context such arbitrariness are expressed by differences between divergent loop integrals with the same degree of divergence
and independent of external momenta with the purpose of bringing about its physical interpretation namely their relation to momentum routing invariance (MRI) in an arbitrary Feynman diagram. Some regularizations may break MRI, an inevitable consequence of energy-momentum conservation at the vertices of Feynman diagrams. The striking connection between momentum routing invariance and preservation of gauge symmetry was realized long ago by t' Hooft and Veltman \cite{'tHooft:1972fi}, by Jackiw in \cite{Treiman:1986ep} as well as by Elias et al. in \cite{Elias:1982sq}. In \cite{Ferreira:2011cv} some of us established the interplay between the vanishing of such arbitrary parameters expressed by surface terms and Abelian gauge invariance in the framework of Implicit Regularization (IReg) . In this four-dimensional method, regularization dependent\\ terms (surface terms) can be extracted out in a consistent way allowing a clear discussion of the ambiguities involved in the manipulation of divergent integrals. Such scheme may be generalized to arbitrary (integer) dimensions and to arbitrary loop order in perturbation theory complying with Lorentz invariance and unitarity as dictated by the local Bogoliubov's $R$-operation based on the BPHZ theorem \cite{Bogoliubov:1957gp,Parasiuk:1960,Hepp:1966eg,Zimmermann:1969jj,Cherchiglia:2010yd}. Therefore, instead of just adding the result of a different method to the literature we intend to show that the discussions presented in \cite{Marciano:2011gm,Piccinini:2011az,Shao:2011wx,Bursa:2011aa,Huang:2011yf} can all be explained using just one framework.

The paper is organized as follows: in section \ref{general} we discuss some regularization dependent integrals and present our parametrizations. Section \ref{Higgs} is dedicated to the calculation of the Higgs decay into two photons in the unitary gauge. In section \ref{Photon} we discuss the result of two photon scattering in the framework of IReg. Finally, section \ref{conclusion} is devoted to our concluding remarks.

\section{A general view of regularization dependent integrals}
\label{general}

In this section we discuss on general grounds the issue of regularization dependent integrals leaving the physical calculations of the Higgs decay as well as of the two photon scattering to subsequent sections. Proceeding this way we hope to set the subject, both from a conceptual and technical point of view, in a consistent and self-contained way allowing a clearer discussion of the examples just cited. 

As is well known perturbative Quantum Field Theoretical calculations involve integration in the momentum loops which must be regularized due to ultraviolet and sometimes infrared divergences. The renormalization program consistently redefines physical degrees of freedom order by order in perturbation theory. Symmetry requirements may either be ensured by an invariant regularization or imposed as constraint equations dictated by Ward-Slavnov-Taylor identities order by order in the loops. Yet a little calculational tedious, the latter procedure is perfectly sound for both anomaly free theories and models in which the quantum symmetry breaking mechanism is well known.

A plethora of regularization schemes have been constructed to be used where gauge invariant DReg may fail, namely in the so called dimensional specific theories among which supersymmetric, chiral and topological quantum field theories figure in. A natural question would be which basic properties should a method that does not resort to analytical continuation in the space-time dimension should retain in order to be invariant. We start by illustrating with  simple examples following \cite{Varin:2006de}. Let $\Delta$ be the superficial degree of divergence of a $1$-loop integral where the momentum $k$ runs. Consider the following $\Delta=2$ integrals,
\be
A = \int_k \frac{k^2}{(k^2-m^2)^2},
\ee
and
\be
B = I_{quad} (m^2) + m^2 I_{log} (m^2),
\ee
where $\int_k \equiv \int d^4k/(2 \pi)^4$ and we recover the standard notation of Implicit Regularization
\be
I_{log} (m^2) \equiv \int_k \frac{1}{(k^2- m^2)^2},
\ee
and
\be
I_{quad} (m^2) \equiv \int_k \frac{1}{(k^2- m^2)}.
\ee
We expect  $A=B$ be guaranteed by any regularization procedure. However this is not the case. Proper-time regularization \cite{ZinnJustin:1993wc}, for instance, introduces a cut-off $\Lambda$ after Wick rotation via the following identity at the level of propagators
\begin{align}
\frac{\Gamma(n)}{(k^2+m^2)^n} = \int_0^{\infty} d\tau \tau^{n-1} e^{- \tau (k^2+m^2)} \rightarrow \int_{1/\Lambda^2}^{\infty}  d\tau \tau^{n-1} e^{- \tau (k^2+m^2)}.
\end{align}
Thus it is trivial to obtain within the proper-time method that $A \neq B$ since
\be
A_{\Lambda} = \frac{-2 i}{(4 \pi)^2} (\Lambda^2 - m^2 \ln \Lambda^2/m^2),
\ee
whereas
\be
B_{\Lambda} = \frac{- i}{(4 \pi)^2} (\Lambda^2 - 2 m^2 \ln \Lambda^2/m^2).
\ee
On the other hand it is straightforward to show that standard DReg leads to $A=B$. To circumvent such discrepancy the authors of \cite{Varin:2006de} define a $n$-dimensional integral
\be
I (\alpha, \beta) = \int_k^{n} \frac{1}{(\alpha k^2 + \beta m^2)},
\ee
with arbitrary $\alpha$ and $\beta$, in order to write
\be
A = - \frac{\partial}{\partial \alpha} I(\alpha,\beta)\Big|_{\alpha=\beta=1, n=4},
\ee
and
\be
B = I(\alpha,\beta)\Big|_{\alpha=\beta=1, n=4}+ \frac{\partial}{\partial \beta} I(\alpha,\beta)\Big|_{\alpha=\beta=1, n=4}.
\ee
Then resorting to proper time regularization one gets
\begin{align}
I(\alpha,\beta)_{\Lambda} = \alpha^{-n/2} \int_k^{n} \frac{1}{(k^2-\beta m^2)} = \frac{-\alpha^{n/2}i}{(4 \pi)^2} (\Lambda^2 - \beta m^2 \ln (\Lambda^2/m^2)),
\end{align}
from which is obtained
\be
A^n_\Lambda = \frac{-i}{(4 \pi)^2} \Big( \frac{n}{2} \Lambda^2 - \frac{n}{2} m^2 \ln (\Lambda^2/m^2)\Big),
\ee
and
\be
B^n_\Lambda = \frac{-i}{(4 \pi)^2} \Big( \Lambda^2 - 2 m^2 \ln (\Lambda^2/m^2)\Big).
\ee
Whilst keeping $n=4$ violates $A=B$, the choices $n=4$ in the term $\propto \ln \Lambda^2 $ and $n=2$ in the term $\propto \Lambda^2 $ lead $A$ to coincide with $B$ at regularized level. Yet arbitrary the authors consider such prescription, which is generalizable to other integrals in Feynman amplitudes, a concrete realization for a  four-dimensional regularization. They claim that Veltman in \cite{Veltman:1980mj} already notices that quadratic divergences are associated with $n=2$ whereas logarithmic divergences have to be treated in $n=4$ in DReg. Other authors have used a similar approach \cite{Harada:2003jx, Harada:2001rf, Harada:2000at}.

Let us now consider another related example. Consider the effect of a shift in the integration variable of a four-dimensional integral. As is well known such shifts  accompany surface terms in more than logarithmically divergent integrals. Their value is highly regularization dependent. For instance take the difference between two linearly divergent integrals for $\omega = 2$
\be
\Delta_1 =\int^{2 \omega}_k \frac{k_\mu}{[(k-p)^2-m^2]^2} - \int_k^{2 \omega} \frac{(k+p)_\mu}{[k^2-m^2]^2}.
\label{Delta1}
\ee
Clearly $\Delta_1 =0$ in DReg because in this method no surface terms accompany shifts in the integration variable. In \cite{Elias:1982sq} the authors generalize the procedure adopted by Jauch and Rohrlich in \cite{Jauch:1955} to evaluate $\Delta_1$ for $\omega$ exactly equal to 2. Their purpose was founded on the physical motivation of constructing four-dimensional regularizations with properties compatible with DReg. By defining
\be
I^{2n+1,r}_{\mu_1 \ldots \mu_{2n+1}} = \int^{2 \omega}_k \frac{\prod_{j=1}^{2n+1}k_{\mu_j}}{[(k-p)^2-m^2]^r},
\ee
and
\be
J^{2n+1,r}_{\mu_1 \ldots \mu_{2n+1}} = \int^{2 \omega}_k \frac{\prod_{j=1}^{2n+1}(k+p)_{\mu_j}}{[k^2-m^2]^r},
\ee
it is shown in \cite{Elias:1982sq} that whilst $I=J$ for $2 \omega + 2 n + 1 - 2 r < 1$, if $2 > 2 \omega + 2 n + 1 - 2 r > 1$ then
\be
I^{2n+1,r}_{\mu_1 \ldots \mu_{2n+1}} - J^{2n+1,r}_{\mu_1 \ldots \mu_{2n+1}} = \frac{-i (2\pi)^4\pi^\omega G_{n,2n+1}(p)}{\Gamma (\omega)} \delta_{r,\omega+n},
\ee
with
\be
G_{n,2n+1}(p)=\frac{g_{\mu_{j_1}\mu_{j_2}} \ldots
g_{\mu_{j_{2n-1}}\mu_{j_{2n}}}p_{\mu_{j_{2n+1}}}
\sigma^{j_1 \ldots j_{2n+1}}}{\Gamma (\omega)^{-1} \Gamma (\omega + n
+ 1)n ! 2^{2n}},
\ee
and
\be
\sigma^{j_1 \ldots j_{2n+1}} = \epsilon^{j_1 \ldots j_{2n+1}} (-)^{sign(\epsilon)}.
\ee
For $n=0$ we immediately obtain
\be
\Delta_1 = \frac{-i \pi^2 (2 \pi)^4}{2} \delta_{\omega 2} p_\mu.
\ee
A similar expression may be obtained for more than linearly divergent variable shifted integrals. It is immediate from above that the Kronecker delta signs a discontinuity in the dimensionality $\omega$. The authors use these results to back up an integer dimensional regularization called Preregularization where the freedom of momentum routing in the loops is chosen to cancel out some surface terms thus preserving Ward identities in chiral anomalies or supersymmetry \cite{Elias:1983jz, Elias:1986ra, Elias:1984zs}. A relevant question, given that  shifts of integration variables are regularization dependent, would be to verify whether the argument could be turned the other way around, namely to exploit the consequences of momentum routing invariance over regularization schemes. Some technicalities deserve attention. Symmetric integration in $n$ (integer) dimensions, namely
$k_\mu k_\nu \rightarrow g_{\mu \nu} k^2/n$ under integration in $k$ for divergent integrals does $not$ hold in general and has been a source of  disagreements in loop calculations as well discussed in \cite{PerezVictoria:2001ej} in the context of CPT violation in quantum field theory and used in \cite{Gastmans:2011wh} to study  the Higgs decay into two photons. In particular symmetric integration was used in  \cite{Jauch:1955} to evaluate $\Delta_1$.

We proceed to write down a general parametrization for loop integrals which incorporates explicitly arbitrary constants which will be fixed on physical grounds. Later on we propose an alternative description of ultraviolet (and infrared) divergences in terms of basic divergent integrals. In such description undetermined regularization dependent constants are expressed in terms of a special set of well known surface terms, namely integrals of total divergences in momentum space, whose contact with momentum routing invariance in the diagrams is immediate as well is their generalization to arbitrary loop order. In order to isolate the basic loop integrals from Feynman amplitudes an identity at the level of the integrand,
\begin{align}
 \frac{1}{[(k+p)^2-m^2]} = \sum_{j=0}^{N} \frac{(-1)^j (p^2+ 2 p
\cdot k)^j}{(k^2-m^2)^{j+1}} +
\frac{(-1)^{N+1} (p^2 + 2 p\cdot k)^{N+1}}{(k^2 -m^2)^{N+1}
[(k+p)^2-m^2]} \, ,
\end{align}
can be judiciously used to extract external momentum dependence from loop integrals. Such operation at the level of integrands somewhat resembles the renormalization procedure originally proposed by Bogoliubov, Parasiuk, Hepp and Zimmermann (BPHZ) \cite{Bogoliubov:1957gp,Parasiuk:1960,Hepp:1966eg,Zimmermann:1969jj} in which divergent Green functions are Taylor expanded up to the order needed to reach convergent integrals.  We assume an implicit regulator under the integration sign which acts on the physical dimension of the underlying theory avoiding conflicts with space-time and internal algebras sensitive to dimensional continuation. Consider the derivative of $I_{log} (m^2)$ in $d$ (integer) space-time dimensions,
\bq
\frac{d I_{log}(m^2)}{d m^{2}}&=&-\frac{b_{d}}{m^{2}},\nonumber\\
\frac{d I_{log}^{\mu \nu}(m^2)}{d m^{2}}&=&- \frac{g^{\mu \nu}}{d}\frac{b_{d}}{m^{2}},
\eq
where for future reference
\be
b_d = \frac{i}{(4 \pi)^{d/2}}\frac{(-)^{d/2}}{\Gamma (d/2)}.
\ee
A general parametrization which obeys the relations above is given by
\begin{align}
I_{log} (m^2)&=  b_{d}\ln\left(\frac{\Lambda^2}{m^2}\right) + \alpha_1,\nonumber\\
I_{log}^{\mu\nu}(m^2)&= \frac{g^{\mu\nu}}{d}\Bigg[b_{d}\ln\left(\frac{\Lambda^2}{m^2}\right) + \alpha_1'\Bigg],
\label{resilog}
\end{align}
\noindent
where $\alpha_{1}$, $\alpha'_{1}$ are arbitrary dimensionless regularization dependent constants, $\Lambda$ is an ultraviolet cut-off,
and
\be
I_{log}^{\mu\nu}(m^2) = \int_k \frac{k^\mu k^\nu}{(k^2- m^2)^3}.
\ee
\noindent
In a similar fashion
\begin{align}
\frac{d I_{quad} (m^2)}{d m^{2}}&=\frac{(d-2)}{2}\,I_{log}(m^2),\nonumber\\
\frac{d I_{quad}^{\mu\nu}(m^2)}{d m^{2}}&= \left(\frac{d}{2}\right) I_{log}^{\mu\nu}(m^2),
\end{align}
\noindent
where the expression for $I_{quad}^{\mu\nu}(m^2)$ is now clear, namely a basic quadratically divergent integral containing two loop momenta with Lorentz indices $\mu$ and $\nu$.  Again, a parametrization that complies with the relations above is
\small
\begin{align}
I_{quad}(m^2)=&\frac{(d-2)}{2}\Bigg[\alpha_{2}\Lambda^{2}+ b_{d}m^{2}\ln\left(\frac{\Lambda^2}{m^2}\right)+ \alpha_3 m^{2}\Bigg],\nonumber\\
I_{quad}^{\mu\nu}(m^2)=&\frac{g^{\mu\nu}}{2}\Bigg[\alpha'_{2}\Lambda^{2}+ b_{d}m^{2}\ln\left(\frac{\Lambda^2}{m^2}\right)\
+ \alpha_3' m^2\Bigg],
\label{resiquad}
\end{align}
\normalsize
in which all regularization dependence is encoded in the $\alpha$'s. Some comments are in order. It is economical and neat to write basic divergent integrals  in terms of $\{ I_{log} (m^2), I_{quad} (m^2) \ldots \}$ without Lorentz indices in internal momenta, in other words expressing $I_{log}^{\mu\nu}(m^2)$ and $I_{quad}^{\mu\nu}(m^2)$, etc. in terms of $I_{log} (m^2)$ and  $I_{quad} (m^2)$ respectively both without resorting to symmetric integration and in a regularization independent way through surface terms. For instance it is straightforward to show that
\begin{align}
\Upsilon_{0}^{\mu\nu} \equiv \int^d_k \frac{\partial}{\partial k_{\mu}}\frac{k^{\nu}}{(k^{2}-m^{2})^{\frac{d}{2}}} = d\Bigg[\frac{g^{\mu\nu}}{d}I_{log}(m^2)-I_{log}^{\mu\nu}(m^2)\Bigg],
\end{align}
\noindent
and
\begin{align}
\Upsilon_{2}^{\mu\nu} \equiv \int^d_k\frac{\partial}{\partial k_{\mu}}\frac{k^{\nu}}{(k^{2}-m^{2})^{\frac{d-2}{2}}} = (d-2)\Bigg[\frac{g^{\mu\nu}}{(d-2)}I_{quad}(m^2)-I_{quad}^{\mu\nu}(m^2)\Bigg].
\end{align}

The surface terms $\Upsilon$'s are regularization dependent terms which however can be shown to be physical meaningful and therefore be fixed. That is because although the intrinsic (regularization dependent) parameters in loop integrals are indeed ambiguous, the well adjusted relation between them expressed by the $\Upsilon$'s are not. In other words in the process of reducing the set of loop integrals to basic divergent integrals it can be shown that the vanishing of surface terms expressed by the $\Upsilon$'s reflects momentum routing invariance in the loops of a Feynman diagram \cite{Ferreira:2011cv, Battistel:1998sz}. Attributing spurious values to such surface terms is the root of quantum symmetry breaking by regularizations. Once we attach a physical meaning to them, as it is proposed in the Implicit Regularization program we may regularize infinities in a regularization independent fashion because the renormalization constants can be defined in terms of basic divergent integrals themselves. To see that they are regularization dependent we can use the parametrizations (\ref{resilog}) and (\ref{resiquad}) to obtain
\be
\Upsilon_0^{\mu \nu} \propto g^{\mu \nu} (\alpha_1 - \alpha_1'),
\ee
and
\be
\Upsilon_2^{\mu \nu} \propto g^{\mu \nu} [(\alpha_2 - \alpha_2')\Lambda^2 + (\alpha_3 - \alpha_3')m^2].
\ee
For instance in the four-dimensional case $\Upsilon_0^{\mu \nu}=g^{\mu \nu} [i/8(4\pi)^2]$ and $\Upsilon_2^{\mu \nu}=  g^{\mu \nu}\Lambda^2 [i/4(4\pi)^2]$ in sharp cut-off regularization \cite{Felipe:2011rs} whilst they are both zero in DReg. As for the examples we presented earlier, it is immediate that $A=B$ within our approach because summing and subtracting $m^2$ in the numerator of $A$ leads to $B$. Whenever even powers of internal momenta appear in the numerator, one can always make use of such artifice to avoid ambiguous symmetric integration \cite{Pontes:2007fg}. As for $\Delta_1$ in equation (\ref{Delta1}) one obtains within Implicit Regularization
\be
\Delta_1^{IR} = \Upsilon_0^{\mu \nu} p_\nu.
\ee
In \cite{Ferreira:2011cv} we demonstrate that momentum routing invariance in Feynman diagrams (enforced by setting all surface terms $\Upsilon$' s to zero) leads automatically to Abelian gauge invariance at arbitrary loop order.

For the sake of completeness we draw a few remarks regarding renormalization and generalization to arbitrary loop order in four space-time dimensions within this approach. To define a mass independent scheme we use the regularization independent relation
\begin{equation}
I_{log}(m^{2}) = I_{log}(\lambda^{2}) + b\ln\left(\frac{\lambda^{2}}{m^{2}}\right),
\end{equation}
where $\lambda \ne 0$ plays the role of renormalization group scale (see \cite{Ferreira:2011cv} and references therein). After subtraction of subdivergences according to BPHZ formalism we may define the divergence of $n^{th}$ loop order in terms of basic divergent integrals for both massive and massless theories \cite{Cherchiglia:2010yd} in the form
\be
I_{log}^{(n)}(m^{2}) \equiv \int_{k}\frac{1}{(k^{2}-m^{2})^{2}}\ln^{n-1}\left(-\frac{(k^{2}-m^{2})}{\lambda^{2}}\right),
\ee
which obeys
\be
I_{log}^{(n+1)}(m^{2}) = I_{log}^{(n+1)}(\lambda^{2}) - b\sum_{i=1}^{n+1}\frac{n!}{i!}\ln^{i}\left(\frac{m^{2}}{\lambda^{2}}\right).
\ee
Likewise
\begin{align}
\frac{d I_{log}^{(n)}(\lambda^2)}{d
\lambda^{2}}&=-\frac{(n-1)}{\lambda^{2}}I_{log}^{(n-1)}(\lambda^2)+\frac{b_{d}}{\lambda^{2}}A^{(n)},\nonumber\\
\frac{d I_{log}^{(n)\,\mu\nu}(\lambda^2)}{d
\lambda^{2}}&=-\frac{(n-1)}{\lambda^{2}}I_{log}^{(n-1)\,\mu\nu}(\lambda^2)+\frac{g^{\mu\nu}}{2}\frac{b_{d}}{\lambda^{2}}B^{(n)}.
\label{gerder}
\end{align}
\noindent
After some algebra one can demonstrate that the parametrization below respects (\ref{gerder}),
\small
\begin{align}
I_{log}^{(n)}(\lambda^2)=\sum\limits_{i=1}^{n}&\frac{(n-1)!}{(i-1)!}\!\Bigg[\!\frac{(-b_{d})A^{(i)}}{(n-i+1)!}\ln^{n-i+1}\!\!\left(\frac{\Lambda^{2}}{\lambda^{2}}\right)+\sum\limits_{j=0}^{n-i}\frac{a_{n-j-i+1}}{j!(n-j-i)!}\ln^{j}\!\!\left(\frac{\Lambda^{2}}{\lambda^{2}}\right)\!\!\Bigg],\nonumber
\end{align}
\begin{align}
I_{log}^{(n)\,\mu\nu}(\lambda^2)=\frac{g^{\mu\nu}}{2}\!\sum\limits_{i=1}^{n}&\frac{(n-1)!}{(i-1)!}\!\Bigg[\!\frac{(-b_{d})B^{(i)}}{(n-i+1)!}\ln^{n-i+1}\!\!\left(\frac{\Lambda^{2}}{\lambda^{2}}\right)+\sum\limits_{j=0}^{n-i}\frac{a'_{n-j-i+1}}{j!(n-j-i)!}\ln^{j}\!\!\left(\frac{\Lambda^{2}}{\lambda^{2}}\right)\!\!\Bigg],
\end{align}
\noindent
\normalsize
where
\small
\begin{align}
A^{(i)}\equiv\Gamma(d/2)\lim_{\delta\rightarrow0}\Bigg[&-(n-1)\sum\limits_{l=0}^{n-2}\binom{n-2}{l}\frac{(-1)^{1+l}}{\delta^{n-2}}\frac{\Gamma(1-\delta(n-2-l))}{\Gamma(d/2+1-\delta(n-2-l))}+\nonumber\\&+\times\left(\frac{d}{2}\right)\sum\limits_{l=0}^{n-1}\binom{n-1}{l}\frac{(-1)^{1+l}}{\delta^{n-1}}\frac{\Gamma(1-\delta(n-1-l))}{\Gamma(d/2+1-\delta(n-1-l))}\Bigg],\nonumber\\
B^{(i)}\equiv\Gamma(d/2)\lim_{\delta\rightarrow0}\Bigg[&-(n-1)\sum\limits_{l=0}^{n-2}\binom{n-2}{l}\frac{(-1)^{1+l}}{\delta^{n-2}}\frac{\Gamma(1-\delta(n-2-l))}{\Gamma(d/2+2-\delta(n-2-l))}+\nonumber\\&+\left(\frac{d+2}{2}\right)\sum\limits_{l=0}^{n-1}\binom{n-1}{l}\frac{(-1)^{1+l}}{\delta^{n-1}}\frac{\Gamma(1-\delta(n-1-l))}{\Gamma(d/2+2-\delta(n-1-l))}\Bigg],
\end{align}
\noindent
\normalsize
and $a_{i}$, $a'_{i}$ are arbitrary constants. The surface terms read
\begin{align}
\frac{1}{2}\sum_{j=1}^{n}\left(\frac{2}{d}\right)^j&\frac{(n-1)!}{(n-j)!}\Upsilon_{0}^{(n)\mu\nu}=-I_{log}^{(n)\,\mu
\nu}(\lambda^2)+\frac{g^{\mu \nu}}{2}\sum_{j=1}^{n}\left(\frac{2}{d}\right)^j\!\frac{(n-1)!}{(n-j)!}I_{log}^{(l-j+1)}(\lambda^2).
\end{align}
\noindent
Generalization to an arbitrary number of Lorentz indices is equally straightforward.

\section{Higgs decay into two photons}
\label{Higgs}

\begin{figure}[t]
\begin{center}
\includegraphics[scale=0.56]{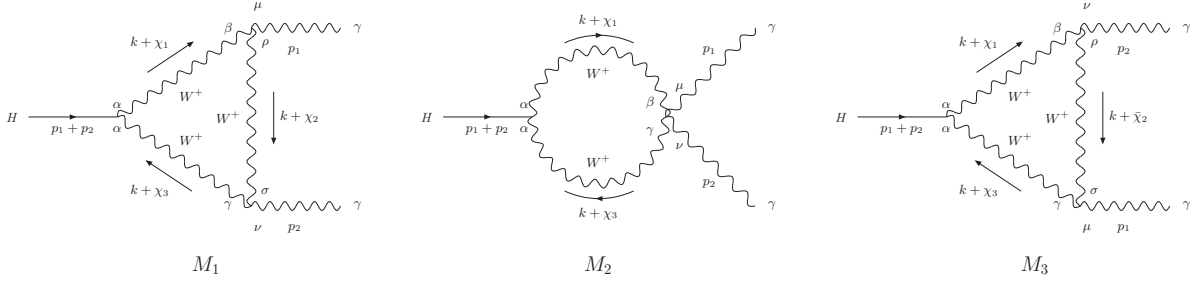}
\end{center}
\vspace{-1.0cm}
\caption{Diagrams with arbitrary momentum routing $\chi$}
\label{diagrams}
\end{figure}

In this section we will study the W loop contributions to the Higgs decay into two photons. Using the unitary gauge we have only three Feynman diagrams to evaluate (fig. \ref{diagrams}). Notice that we are not choosing a specific routing for the diagrams since we want to study how the final amplitude depends on it.

The sum of the three diagrams can be simplified to the expression (Feynman rules as well as the basic steps to arrive at the equations below are presented on Appendix \ref{calculations}) \footnote{We define $q_{i}=k+\chi_{i}$, $\bar{q}_{i}=k+\bar{\chi}_{i}$, $\int\limits_{k}=\int\frac{d^4k}{(2\pi)^{4}}$ and use relations $p_{i}^{2}=0$ and $(p_1 + p_2)^2 = M_h^2$}
\begin{align}
M&=ie^{2}gM_{w}\Big[M_{\mu\nu}^{(a)}+M_{\mu\nu}^{(b)}+M_{\mu\nu}^{(c)}\Big]({\epsilon_{1}}^{\mu})^{\ast}({\epsilon_{2}}^{\nu})^{\ast}+(p_{1}\leftrightarrow p_{2},\mu\leftrightarrow\nu),\\
M_{\mu\nu}^{(a)}&=-\frac{4}{M^{2}_{w}}\Big[g_{\mu\nu}(p_{1})^{\alpha} (p_{2})^{\beta}I^{(3)}_{\alpha\beta}+(p_{1}\cdot p_{2})I^{(3)}_{\mu\nu}-(p_{1})_{\nu}(p_{2})^{\alpha}I^{(3)}_{\mu\alpha}-(p_{2})_{\mu}(p_{1})^{\alpha}I^{(3)}_{\nu\alpha}\Big]+\nonumber\\&\quad+\frac{2}{M^{2}_{w}}\Big[g_{\mu\nu}(p_{1}\cdot p_{2})-(p_{2})_{\mu}(p_{1})_{\nu}\Big]I^{(3)}_{2},\\
M_{\mu\nu}^{(b)}&=\int\limits_{k}\frac{3(g_{\mu\nu}k^{2}-4k_{\mu}k_{\nu})}{(q_{1}^{2}-M^{2}_{w})(q_{2}^{2}-M^{2}_{w})(q_{3}^{2}-M^{2}_{w})},\\
M_{\mu\nu}^{(c)}&=6g_{\mu\nu}\Big[(p_{1}\cdot p_{2})I^{(3)}_{0}-(p_{1})^{\alpha}I^{(3)}_{\alpha}-\frac{M^{2}_{w}}{2}I^{(3)}_{0}\Big]+6\Big[2(p_{1})_{\nu}I^{(3)}_{\mu}-(p_{2})_{\mu}(p_{1})_{\nu}I^{(3)}_{0}\Big],\\
\label{Before}
I^{(3)}_{0,2,\mu,\mu\nu}&=\int\limits_{k}\frac{1,k^2,k_{\mu},k_{\mu}k_{\nu}}{(q_{1}^{2}-M^{2}_{w})(q_{2}^{2}-M^{2}_{w})(q_{3}^{2}-M^{2}_{w})}.
\end{align}

As one may notice only $M_{\mu\nu}^{(a)}$ and $M_{\mu\nu}^{(b)}$ contain divergent terms. At this point we must choose a regularization in order to deal properly with such terms. We employ IReg which allows us to express divergent integrals in terms of loop momenta only (for a review see \cite{Ferreira:2011cv} and references therein). A characteristic of IReg is that all regularization-dependent objects (surface terms) can be consistently treated allowing a clear discussion about ambiguities as will be seen below. Explicitly they are given by
\begin{align}
\Upsilon_0^{\mu \nu}=g^{\mu\nu}\Gamma_{0}=\int\limits_{k}\frac{\partial}{\partial k_{\mu}}\frac{k^{\nu}}{(k^2-M^{2}_{w})^{2}}=\int\limits_{k}\frac{g^{\mu\nu}}{(k^2-M^{2}_{w})^{2}}-4\int\limits_{k}\frac{k^{\mu}k^{\nu}}{(k^2-M^{2}_{w})^{3}}.
\label{TS}
\end{align}
Therefore the first term can be expressed as
\begin{align}
M_{\mu\nu}^{(a)}=\frac{\big[(p_{2})_{\mu}(p_{1})_{\nu}-g_{\mu\nu}(p_{1}\cdot p_{2})\big]}{M^{2}_{w}}\Bigg[\frac{i}{16\pi^2}-2\Gamma_{0}\Bigg].
\label{Ma}
\end{align}

The first point to be noticed is that this term is gauge invariant and, in general, ambiguous since it depends on a surface term. Another feature is that it does not depend on\footnote{We define $\tau=\frac{M^{2}_{h}}{4M^{2}_{w}}$} $\tau$ which give us a clue that it may be the term missing on \cite{Gastmans:2011wh}. In fact, if one performs a symmetric regularization in four-dimensions (by replacing $k_{\mu}k_{\nu}\rightarrow g_{\mu\nu}k^2/4$) it will be null. In other words a four-dimensional regularization that resorts to such substitution evaluates the surface term to a precise value, in this case $i/32\pi^2$. On the other hand, if one uses DReg the surface term will vanish which furnishes a \textbf{non-null} amplitude in the limit $\tau^{-1}\rightarrow 0$. In the framework of IReg there is no reason \textsl{a priori} to favor one of these two values since we are dealing with  ambiguous objects in nature. From our perspective physical conditions, other than the regularization method, should constrain the value the surface term should assume. In general, one such condition is to impose gauge invariance, however, since this term is \textbf{already} gauge invariant, this consideration will not fix it. Therefore, we should leave it arbitrary and proceed with the calculation of the amplitude for now. The sum of the two last terms is\footnote{Where we define \begin{equation}
f(\tau)=\left\{\begin{array}{lcc}
\arcsin^2(\sqrt{\tau})&\mbox{for}&\tau\leq 1\,,\\[5mm]
-{\displaystyle\frac{1}{4}\,\left[\ln\frac{1+\sqrt{1-\tau^{-1}}}{1-\sqrt{1-\tau^{-1}}}-i\pi\right]^2}
&\mbox{for}&\tau>1\,.\end{array}\right.\nonumber
\end{equation}}
\begin{align}
M_{\mu\nu}^{(b)}+M_{\mu\nu}^{(c)}&=\frac{i}{16\pi^2M^{2}_{w}}\big[(p_{2})_{\mu}(p_{1})_{\nu}-g_{\mu\nu}(p_{1}\cdot p_{2})\big]\Bigg[\frac{3\tau^{-1}}{2}+\frac{3(2\tau^{-1}-\tau^{-2})f(\tau)}{2}\Bigg]\nonumber\\
&+g_{\mu\nu}(p_{1}\cdot p_{2})\Bigg(\frac{3\tau^{-1}}{2M^{2}_{w}}\Gamma_{0}\Bigg).
\label{Mb}
\end{align}

Readily one may notice the appearance of another surface term due to $M_{\mu\nu}^{(b)}$ which explicitly breaks gauge invariance. Since there are no other terms to consider, one should impose gauge invariance as a physical condition that the whole amplitude should fulfill. Thus the otherwise arbitrary surface term must assume a precise value which in our case is null. This choice also fixes the surface term appearing in (\ref{Ma}) since in the framework of IReg there is no distinction between surface terms coming from integrals with the same degree of divergence and the same Lorentz structure. This approach is different from the one found in \cite{Piccinini:2011az} where a cutoff scheme is used and the ambiguities are parametrized by different boundary conditions for the integrals  appearing in (\ref{Ma}) and (\ref{Mb}). Since these authors consider that each integral is arbitrary and unrelated to each other, they conclude that the imposition of

\vspace{1.7cm}
\noindent
gauge invariance is not enough to give an unambiguously result.

After all these considerations we obtain the amplitude for the Higgs decay into two photons in the framework of IReg
\begin{align}
&M=-\frac{e^2g}{16\pi^2M_{w}}\big[(p_{2})_{\mu}(p_{1})_{\nu}-g_{\mu\nu}(p_{1}\cdot p_{2})\big]\big[2+3\tau^{-1}+3(2\tau^{-1}-\tau^{-2})f(\tau)\big]({\epsilon_{1}}^{\mu})^{\ast}({\epsilon_{2}}^{\nu})^{\ast},
\label{result}
\end{align}
\noindent
which agrees with previous ones found in the literature \cite{Ellis:1975ap,Ioffe:1976sd,Shifman:1979eb}.

In the time this work was been written another paper devoted to this decay appeared \cite{Dedes:2012hf}. Their authors have a point of view similar to ours in the sense that ambiguities should be fixed on physical grounds \footnote{It should be emphasized that their definition of the ambiguity is more closely related to the one found in \cite{Jackiw:1999qq}. We, on the other hand, define it by (\ref{TS}) which is more closely related to the preservation of Abelian gauge invariance \cite{Ferreira:2011cv}.}. They use the equivalence theorem as well as the conservation of charge as inputs that their amplitude must fulfill. Since these are consequences of gauge invariance there is no surprise that just the imposition of such requirement gives us an unambiguous result.

Another interesting point discussed there is the role played by momentum routing freedom. From their point of view the loop-momentum of the three diagrams must be chosen in a particular way as to reduce the superficial degree of divergence of the amplitude to a logarithmic one.\footnote{They find that all three diagrams must contain the same momentum routing. Therefore it is no surprise that our result before regularization (\ref{Before}) contains at most logarithmic divergent integrals since we also adopted the same momentum routing for all three diagrams ($\chi_{1}$).} However, from our point of view momentum routing invariance (MRI) is a symmetry that must be respected since it is connected with Abelian gauge invariance as well as supersymmetry preservation \cite{Ferreira:2011cv}. The importance of this statement is particular clear if, instead of considering the calculation of the whole amplitude, one evaluates each diagram \textbf{individually}. Following the reasoning of \cite{Ferreira:2011cv} one finds out that momentum routing dependent terms will arise always multiplied by arbitrary-valued objects (surface terms). Therefore, since individual diagrams are not supposed to be gauge invariant, the only symmetry left in order to fix the ambiguities is demanding momentum routing invariance. As
can be seen, we could have adopted this approach since the beginning of our work avoiding completely the discussion of gauge invariance (since the two symmetries are connected it is not a surprise that the surface terms must be null in both cases). However, in order to make

\vspace{1.0cm}
\noindent
contact with the literature we performed the calculation of the whole amplitude with the same routing for all three diagrams which evidently is not the more general situation. Therefore, it is not a surprise that our result is independent of the momentum routing $\chi_{1}$ even though we still have an ambiguity expressed by $\Gamma_{0}$.

\section{Two photon scattering}
\label{Photon}

In this brief section we would like to comment on the result found in \cite{Kanda:2011vu}. As in the case just analyzed, the problem lies on divergent integrals which appear as intermediate steps of the calculation. Explicitly we have \cite{Liang:2011sj}
\begin{align}
A^{\mu\nu\rho\sigma}&=\int\limits_{k}\frac{m^{4}S_{1}^{\mu\nu\rho\sigma}+2m^{2}\left(2S_{2}^{\mu\nu\rho\sigma}-k^{2}S_{1}^{\mu\nu\rho\sigma}\right)}{\left(k^{2}-m^{2}\right)^{4}}\nonumber\\&+\int\limits_{k}\frac{24k^{\mu}k^{\nu}k^{\rho}k^{\sigma}+\left(k^{2}\right)^{2}S_{1}^{\mu\nu\rho\sigma}-4k^{2}S_{2}^{\mu\nu\rho\sigma}}{\left(k^{2}-m^{2}\right)^{4}},\,
\end{align}
\noindent
where
\begin{align*}
S_{1}^{\mu\nu\rho\sigma} =& g^{\mu\nu}g^{\rho\sigma}+g^{\mu\rho}g^{\nu\sigma}+g^{\mu\sigma}g^{\rho\nu},\\
S_{2}^{\mu\nu\rho\sigma} =& g^{\mu\nu}k^{\rho}k^{\sigma}+g^{\mu\rho}k^{\nu}k^{\sigma}+g^{\mu\sigma}k^{\rho}k^{\nu}+g^{\rho\nu}k^{\mu}k^{\sigma}+g^{\sigma\nu}k^{\rho}k^{\mu}+g^{\rho\sigma}k^{\mu}k^{\nu}.
\end{align*}

As can be readily seen, the integral above is divergent, thus ambiguous. Such statement is particularly clear in the framework of IReg since it is evaluated to $(\Gamma_{0}^{(2)}-4\Gamma_{0}^{(4)})S_{1}^{\mu\nu\rho\sigma}$ where $\Gamma_{0}^{(i)}$ is a surface term coming from an integral with Lorentz structure $k^{\nu_{1}}\cdots k^{\nu_{i}}$. Therefore, there is no preferred value this integral should assume, it should be left arbitrary being fixed by the imposition of physical conditions. As discussed in \cite{Liang:2011sj}, a non-null value for $A^{\mu\nu\rho\sigma}$ implies the breaking of gauge invariance which means the surface terms must obey $\Gamma_{0}^{(2)}=4\Gamma_{0}^{(4)}$ in order to respect such symmetry. Thus, there is no ambiguity left on the final amplitude which as expected agrees with previous results found in the literature \cite{Liang:2011sj}. In summary, as in the case of \cite{Gastmans:2011wh}, the authors of \cite{Kanda:2011vu} performed a symmetric regularization on the integral above which in turn gave a precise value to the surface terms ($A^{\mu\nu\rho\sigma}=(i/96\pi^2)S_{1}^{\mu\nu\rho\sigma}$). Such choice resulted in a different cross section for the two photon scattering than the one found previously in the literature \cite{Karplus:1950zz,Karplus:1950zza}. However, since the integral is ambiguous in nature there is no reason to assume a precise value for the surface terms which must be fixed on physical grounds.

\section{Concluding remarks}
\label{conclusion}

In this work we studied the decay of the Higgs boson into two photons as well as the two photon scattering amplitude. Both processes must have only finite corrections since the photon does not couple with the Higgs boson neither with itself. However, in the intermediate steps of the calculation one may encounter divergent integrals and the issue of regularization is particularly important in order to give a meaningful result. To discuss the ambiguities inherent in such processes we used the framework of Implicit Regularization which can consistently separate the divergent, finite and ambiguous part of any integral. We found out that although the divergent parts cancel as expected there are some ambiguities left (parametrized as surface terms). These should not be fixed by the regularization scheme \textit{a priori}, but should be left arbitrary been determined by physical conditions. In the cases studied here, the condition used was the gauge invariance of the final result which univocally fixed the surface terms thus recovering the amplitude for the Higgs decay as well as the cross section of the two photon scattering found previously in the literature.

\acknowledgments

The authors would like to thank R. Jackiw for helpful discussions. L. A. Cabral would like to thanks UFMG staff for their support during his stay in the institution. A. L. Cherchiglia and Marcos Sampaio acknowledge financial support by FAPEMIG. Marcos Sampaio also acknowledges financial support by CNPq. 

\appendix
\section{Some explicit calculations of the amplitude $H\rightarrow\gamma\gamma$}
\label{calculations}

In this appendix we will show how the terms presented on the calculation of the Higgs decay into two photons can be simplified. We start defining the Feynman rules

\begin{figure}[ht!]
\begin{center}
\includegraphics[scale=0.9]{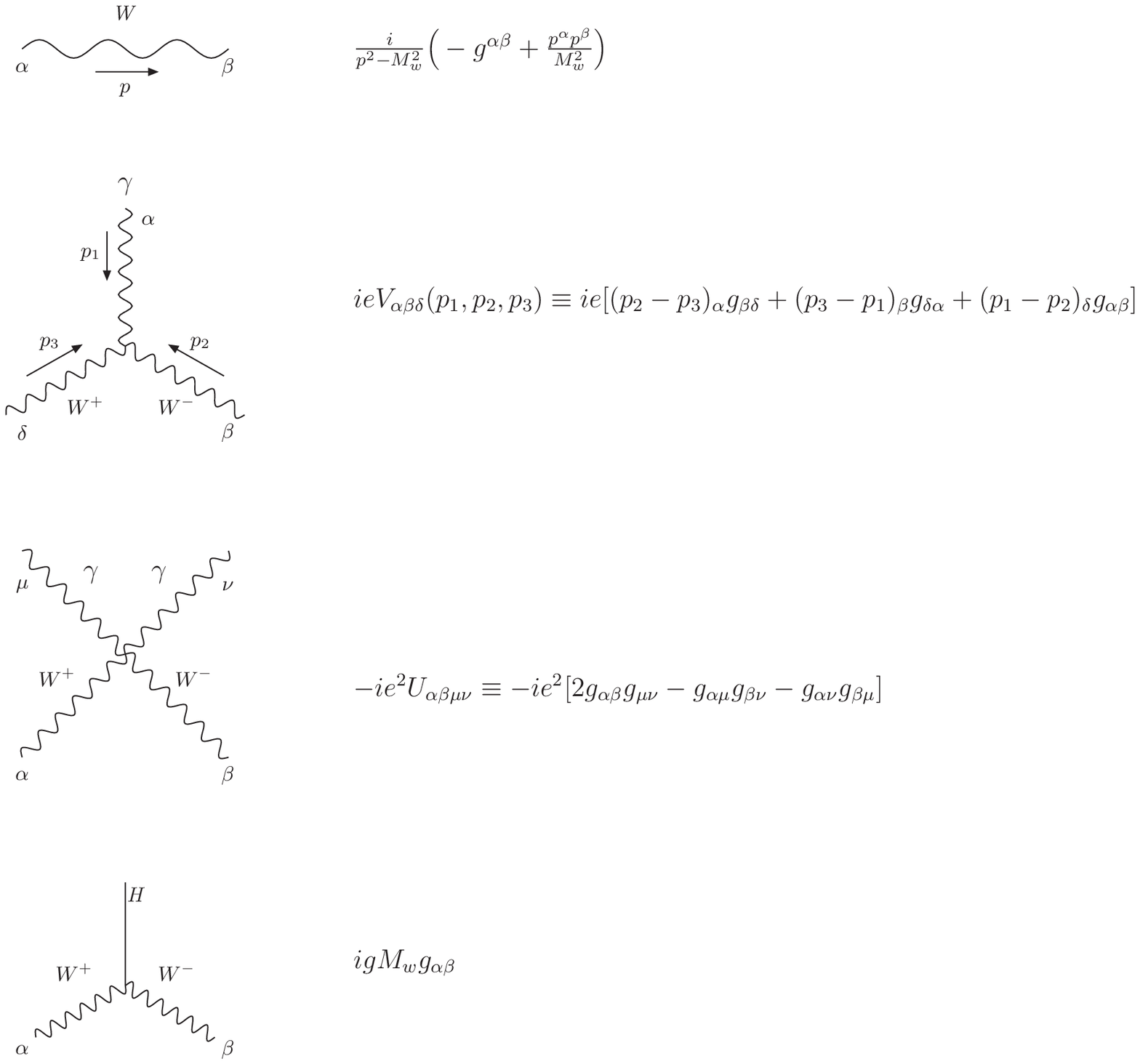}
\end{center}
\vspace{-0.5cm}
\caption{\label{figvertex} Feynman rules.}
\end{figure}
\vspace{3.7cm}

Defining $q_{i}=k+x_{i}$, $\bar{q}_{i}=k+\bar{x}_{i}$ and $\int\limits_{k}=\int\frac{d^4k}{(2\pi)^{4}}$ the diagrams of fig. \ref{diagrams} are expressed as

\begin{align}
M_{1}=ie^{2}gM_{w}\int\limits_{k}&\frac{1}{q_{1}^{2}-M^{2}_{w}}\Big(-g^{\alpha\beta}+\frac{q_{1}^{\alpha}q_{1}^{\beta}}{M^{2}_{w}}\Big)V_{\mu\rho\beta}(-p_{1},-q_{2},q_{1})\frac{1}{q_{2}^{2}-M^{2}_{w}}\Big(-g^{\rho\sigma}+\frac{q_{2}^{\rho}q_{2}^{\sigma}}{M^{2}_{w}}\Big)\times\nonumber\\&V_{\nu\gamma\sigma}(-p_{2},-q_{3},q_{2})\frac{1}{q_{3}^{2}-M^{2}_{w}}\Big(-g^{\gamma\alpha}+\frac{q_{3}^{\alpha}q_{3}^{\gamma}}{M^{2}_{w}}\Big)({\epsilon_{1}}^{\mu})^{\ast}({\epsilon_{2}}^{\nu})^{\ast},
\end{align}
\begin{align}
M_{2}=ie^{2}gM_{w}\int\limits_{k}&\frac{1}{q_{1}^{2}-M^{2}_{w}}\Big(-g^{\alpha\beta}+\frac{q_{1}^{\alpha}q_{1}^{\beta}}{M^{2}_{w}}\Big)U_{\beta\gamma\mu\nu}\frac{1}{q_{3}^{2}-M^{2}_{w}}\Big(-g^{\gamma\alpha}+\frac{q_{3}^{\alpha}q_{3}^{\gamma}}{M^{2}_{w}}\Big)({\epsilon_{1}}^{\mu})^{\ast}({\epsilon_{2}}^{\nu})^{\ast},
\end{align}
\begin{align}
M_{3}=ie^{2}gM_{w}\int\limits_{k}&\frac{1}{q_{1}^{2}-M^{2}_{w}}\Big(-g^{\alpha\beta}+\frac{q_{1}^{\alpha}q_{1}^{\beta}}{M^{2}_{w}}\Big)V_{\nu\rho\beta}(-p_{2},-\bar{q}_{2},q_{1})\frac{1}{\bar{q}_{2}^{2}-M^{2}_{w}}\Big(-g^{\rho\sigma}+\frac{\bar{q}_{2}^{\rho}\bar{q}_{2}^{\sigma}}{M^{2}_{w}}\Big)\times\nonumber\\&V_{\mu\gamma\sigma}(-p_{1},-q_{3},\bar{q}_{2})\frac{1}{q_{3}^{2}-M^{2}_{w}}\Big(-g^{\gamma\alpha}+\frac{q_{3}^{\alpha}q_{3}^{\gamma}}{M^{2}_{w}}\Big)({\epsilon_{1}}^{\mu})^{\ast}({\epsilon_{2}}^{\nu})^{\ast}.
\end{align}

The strategy now is to classify the terms of the integrand according to their dependence on $M_{w}^{-n}$.

\subsection{Terms $M_{w}^{-6}$}

The term coming from $M_{1}$, which we call $M_{1}^{(-6)}$ is\footnote{In the following we will omit the common factor $ie^{2}gM_{w}$ as well the integral in $k$.}:
\begin{align}
&M_{1}^{(-6)}=\Big(\frac{q_{1}^{\alpha}q_{1}^{\beta}}{M^{2}_{w}}\Big)V_{\mu\rho\beta}(-p_{1},-q_{2},q_{1})\Big(\frac{q_{2}^{\rho}q_{2}^{\sigma}}{M^{2}_{w}}\Big)V_{\nu\gamma\sigma}(-p_{2},-q_{3},q_{2})\Big(\frac{q_{3}^{\alpha}q_{3}^{\gamma}}{M^{2}_{w}}\Big)\frac{({\epsilon_{1}}^{\mu})^{\ast}({\epsilon_{2}}^{\nu})^{\ast}}{D(q_{2}^{2}-M^{2}_{w})}\\&\mbox{where}\quad\frac{1}{D}\equiv\frac{1}{(q_{1}^{2}-M^{2}_{w})}\frac{1}{(q_{1}^{3}-M^{2}_{w})}.
\end{align}

Using that $q_{1}^{\beta}q_{2}^{\rho}V_{\mu\rho\beta}(-p_{1},-q_{2},q_{1})({\epsilon_{1}}^{\mu})^{\ast}=0$ we obtain a null contribution. A similar reasoning can be applied to the term coming from $M_{3}$.

\subsection{Terms $M_{w}^{-4}$}

The diagram $M_{1}$ has two contributions. However, one of them is null (due to identity $q_{1}^{\beta}q_{2}^{\rho}V_{\mu\rho\beta}(-p_{1},-q_{2},q_{1})({\epsilon_{1}}^{\mu})^{\ast}=0$) leaving us with the following term
\begin{align}
M_{1}^{(-4)}=\frac{q_{1}^{\alpha}q_{1}^{\beta}}{M^{2}_{w}}V_{\mu\rho\beta}(-p_{1},-q_{2},q_{1})(-g^{\rho\sigma})V_{\nu\gamma\sigma}(-p_{2},-q_{3},q_{2})\frac{q_{3}^{\alpha}q_{3}^{\gamma}}{M^{2}_{w}}\frac{({\epsilon_{1}}^{\mu})^{\ast}({\epsilon_{2}}^{\nu})^{\ast}}{D(q_{2}^{2}-M^{2}_{w})}.
\end{align}
\noindent
With the help of identity 
\begin{align}
q_{1}^{\beta}V_{\mu\rho\beta}(-p_{1},-q_{2},q_{1})({\epsilon_{1}}^{\mu})^{\ast}=\big\{-(q_{2})_{\mu}(q_{2})_{\rho}+M^{2}_{w}g_{\mu\rho}+[(q_{2})^{2}-M^{2}_{w}]g_{\mu\rho}\big\}({\epsilon_{1}}^{\mu})^{\ast},
\end{align}
\noindent
we can separate $M_{1}^{(-4)}$ into three terms. The first one is null due to identity
\begin{align}
q_{3}^{\gamma}q_{2}^{\sigma}V_{\nu\gamma\sigma}(-p_{2},-q_{3},q_{2})({\epsilon_{2}}^{\nu})^{\ast}=0.
\end{align}
The second is proportional to $M_{w}^{-2}$ and will be treated in the next section (we call it $M_{1}^{(-2;-4)}$). In the third one ($M_{1}^{(-4;2d)}$) we cancel one of the denominators to obtain
\begin{align}
M_{1}^{(-4;2d)}=\frac{q_{1}^{\alpha}}{M^{2}_{w}}g_{\mu\rho}(-g^{\rho\sigma})V_{\nu\gamma\sigma}(-p_{2},-q_{3},q_{2})\frac{q_{3}^{\alpha}q_{3}^{\gamma}}{M^{2}_{w}}\frac{({\epsilon_{1}}^{\mu})^{\ast}({\epsilon_{2}}^{\nu})^{\ast}}{D}.
\end{align}

Performing the exchange $p_{1}\leftrightarrow p_{2}$ and $\mu\leftrightarrow\nu$ we obtain the contributions from diagram $M_{3}$. For diagram $M_{2}$ we have
\begin{align}
M_{2}^{(-4)}=\frac{q_{1}^{\alpha}q_{1}^{\beta}}{M^{2}_{w}}U_{\beta\gamma\mu\nu}\frac{q_{3}^{\alpha}q_{3}^{\gamma}}{M^{2}_{w}}\frac{({\epsilon_{1}}^{\mu})^{\ast}({\epsilon_{2}}^{\nu})^{\ast}}{D}.
\end{align}

Summing $M_{1}^{(-4;2d)}+M_{2}^{(-4)}+M_{3}^{(-4;2d)}$ we obtain a null result.

\subsection{Terms $M_{w}^{-2}$}

The contributions coming from the diagram $M_{1}$ are

\begin{align}
M_{1}^{(-2)}=&\frac{q_{1}^{\alpha}q_{1}^{\beta}}{M^{2}_{w}}V_{\mu\rho\beta}(-p_{1},-q_{2},q_{1})(-g^{\rho\sigma})V_{\nu\gamma\sigma}(-p_{2},-q_{3},q_{2})(-g^{\gamma\alpha})\frac{({\epsilon_{1}}^{\mu})^{\ast}({\epsilon_{2}}^{\nu})^{\ast}}{D(q_{2}^{2}-M^{2}_{w})}\nonumber\\
&+(-g^{\alpha\beta})V_{\mu\rho\beta}(-p_{1},-q_{2},q_{1})\frac{q_{2}^{\rho}q_{2}^{\sigma}}{M^{2}_{w}}V_{\nu\gamma\sigma}(-p_{2},-q_{3},q_{2})(-g^{\gamma\alpha})\frac{({\epsilon_{1}}^{\mu})^{\ast}({\epsilon_{2}}^{\nu})^{\ast}}{D(q_{2}^{2}-M^{2}_{w})}\nonumber\\
&+(-g^{\alpha\beta})V_{\mu\rho\beta}(-p_{1},-q_{2},q_{1})(-g^{\rho\sigma})V_{\nu\gamma\sigma}(-p_{2},-q_{3},q_{2})\frac{q_{3}^{\alpha}q_{3}^{\gamma}}{M^{2}_{w}}\frac{({\epsilon_{1}}^{\mu})^{\ast}({\epsilon_{2}}^{\nu})^{\ast}}{D(q_{2}^{2}-M^{2}_{w})}.
\end{align}

We use identities
\begin{align}
q_{1}^{\beta}V_{\mu\rho\beta}(-p_{1},-q_{2},q_{1})({\epsilon_{1}}^{\mu})^{\ast}=\big\{-(q_{2})_{\mu}(q_{2})_{\rho}+M^{2}_{w}g_{\mu\rho}+[(q_{2})^{2}-M^{2}_{w}]g_{\mu\rho}\big\}({\epsilon_{1}}^{\mu})^{\ast},\\
q_{3}^{\gamma}V_{\nu\gamma\sigma}(-p_{2},-q_{3},q_{2})({\epsilon_{2}}^{\nu})^{\ast}=\big\{-(q_{2})_{\nu}(q_{2})_{\sigma}+M^{2}_{w}g_{\nu\sigma}+[(q_{2})^{2}-M^{2}_{w}]g_{\nu\sigma}\big\}({\epsilon_{2}}^{\nu})^{\ast},
\end{align}
\noindent
in order to separate $M_{1}^{(-2)}$ into three terms: $M_{1}^{(-2;2d)}$ which has only two denominators, $M_{1}^{(0;-2)}$ which is proportional to $M_{w}^{0}$ and $M_{1}^{(-2;3d)}$.

As before, the diagram $M_{3}$ furnishes similar contributions. The diagram $M_{2}$ gives
\begin{align}
M_{2}^{(-2)}=\Bigg[\frac{q_{1}^{\alpha}q_{1}^{\beta}}{M^{2}_{w}}U_{\beta\gamma\mu\nu}(-g^{\gamma\alpha})+(-g^{\alpha\beta})U_{\beta\gamma\mu\nu}\frac{q_{3}^{\alpha}q_{3}^{\gamma}}{M^{2}_{w}}\Bigg]\frac{({\epsilon_{1}}^{\mu})^{\ast}({\epsilon_{2}}^{\nu})^{\ast}}{D}.
\end{align}

Adding $M_{1}^{(-2;2d)}$, $M_{2}^{(-2)}$ and $M_{3}^{(-2;2d)}$ we obtain a null result. Therefore, the remaining terms proportional to $M_{w}^{-2}$ are: $M_{1}^{(-2;3d)}$, $M_{1}^{(-2;-4)}$ and similar contributions from $M_{3}$. Using now the definitions of $q_{i}$ and identities such as $(q_{2})^{2}=[(q_{2})^{2}-M^{2}_{w}]+M^{2}_{w}$, these terms can be simplified to
\begin{align}
M_{1}^{(-2;3d)}&+M_{1}^{(-2;-4)}=M_{1}^{(-2;div)}+M_{1}^{(0;fin)}\nonumber,\\
M_{1}^{(-2;div)}&=\frac{2k^{2}}{M^{2}_{w}}\Big[g_{\mu\nu}(p_{1}\cdot p_{2})-(p_{2})_{\mu}(p_{1})_{\nu}\Big]\frac{({\epsilon_{1}}^{\mu})^{\ast}({\epsilon_{2}}^{\nu})^{\ast}}{D(q_{2}^{2}-M^{2}_{w})}+\nonumber\\&+\frac{4}{M^{2}_{w}}\Big[-g_{\mu\nu}(p_{1})^{\alpha} (p_{2})^{\beta}k_{\alpha}k_{\beta}+(p_{1})_{\nu}(p_{2})^{\alpha}k_{\mu}k_{\alpha}+\nonumber\\&\quad\quad\quad\quad\quad\quad\quad+(p_{2})_{\mu}(p_{1})^{\alpha}k_{\nu}k_{\alpha}-(p_{1}\cdot p_{2})k_{\mu}k_{\nu}\Big]\frac{({\epsilon_{1}}^{\mu})^{\ast}({\epsilon_{2}}^{\nu})^{\ast}}{D(q_{2}^{2}-M^{2}_{w})},\\
M_{1}^{(0;fin)}&=\Big\{[g_{\mu\nu}[-(p_{2})^{\alpha}k_{\alpha}+(p_{1})^{\alpha}k_{\alpha}]-2[-(p_{2})^{\mu}k_{\nu}+(p_{1})^{\nu}k_{\mu}]\Big\}\frac{({\epsilon_{1}}^{\mu})^{\ast}({\epsilon_{2}}^{\nu})^{\ast}}{D(q_{2}^{2}-M^{2}_{w})}.
\end{align}

\subsection{Terms $M_{w}^{0}$}

As before, the terms of order $M_{w}^{0}$ coming from diagram $M_{i}$ will be called $M_{i}^{(0)}$. Therefore, all the terms we have to deal with are summarized below:

\begin{align}
&M_{1}^{(0;-2)}=\Big[(q_{1})^{\gamma}V_{\nu\gamma\mu}(-p_{2},-q_{3},q_{2})+(q_{3})^{\beta}V_{\mu\nu\beta}(-p_{1},-q_{2},q_{1})\Big]\frac{({\epsilon_{1}}^{\mu})^{\ast}({\epsilon_{2}}^{\nu})^{\ast}}{D(q_{2}^{2}-M^{2}_{w})},\\
&M_{1}^{(0;fin)}=\Big\{[g_{\mu\nu}[-(p_{2})^{\alpha}k_{\alpha}+(p_{1})^{\alpha}k_{\alpha}]-2[-(p_{2})^{\mu}k_{\nu}+(p_{1})^{\nu}k_{\mu}]\Big\}\frac{({\epsilon_{1}}^{\mu})^{\ast}({\epsilon_{2}}^{\nu})^{\ast}}{D(q_{2}^{2}-M^{2}_{w})},\\
&M_{1}^{(0)}=(-g^{\alpha\beta})V_{\mu\rho\beta}(-p_{1},-q_{2},q_{1})(-g^{\rho\sigma})V_{\nu\gamma\sigma}(-p_{2},-q_{3},q_{2})(-g^{\gamma\alpha})\frac{({\epsilon_{1}}^{\mu})^{\ast}({\epsilon_{2}}^{\nu})^{\ast}}{D(q_{2}^{2}-M^{2}_{w})},\\
&M_{3}^{(0;-2)}=M_{1}^{(0;-2)}\quad(p_{1}\leftrightarrow p_{2},\mu\leftrightarrow\nu),\\
&M_{3}^{(0;fin)}=M_{1}^{(0;fin)}\quad(p_{1}\leftrightarrow p_{2},\mu\leftrightarrow\nu),\\
&M_{3}^{(0)}=M_{1}^{(0)}\quad(p_{1}\leftrightarrow p_{2},\mu\leftrightarrow\nu),\\
&M_{2}^{(0)}=(-g^{\alpha\beta})U_{\beta\gamma\mu\nu}(-g^{\gamma\alpha})\frac{({\epsilon_{1}}^{\mu})^{\ast}({\epsilon_{2}}^{\nu})^{\ast}}{D}.
\end{align}

The last term can be expressed as
\begin{align}
&M_{2}^{(0)}=M_{2}^{(0;1)}+M_{2}^{(0;3)},\\
&M_{2}^{(0;1)}=\frac{1}{2}(-g^{\alpha\beta})U_{\beta\gamma\mu\nu}(-g^{\gamma\alpha})\frac{(q_{2}^{2}-M^{2}_{w})}{D(q_{2}^{2}-M^{2}_{w})}({\epsilon_{1}}^{\mu})^{\ast}({\epsilon_{2}}^{\nu})^{\ast},\\
&M_{2}^{(0;3)}=M_{2}^{(0;1)}\quad(p_{1}\leftrightarrow p_{2},\mu\leftrightarrow\nu).
\end{align}

Adding $M_{1}^{(0;-2)}$, $M_{1}^{(0;fin)}$, $M_{1}^{(0)}$ and $M_{2}^{(0;1)}$ together we obtain
\begin{align}
&M_{1}^{(0;-2)}+M_{1}^{(0;fin)}+M_{1}^{(0)}+M_{2}^{(0;1)}=M_{t}^{(0,div)}+M_{t}^{(0,fin)},\\
&M_{t}^{(0,div)}=\Big[3(g_{\mu\nu}k^{2}-4k_{\mu}k_{\nu})\Big]\frac{({\epsilon_{1}}^{\mu})^{\ast}({\epsilon_{2}}^{\nu})^{\ast}}{D(q_{2}^{2}-M^{2}_{w})},\\
&M_{t}^{(0,fin)}=3g_{\mu\nu}\Big[2(p_{1}\cdot p_{2}) -2(p_{1})^{\alpha}k_{\alpha}-M^{2}_{w}\Big]+6\Big[2(p_{1})_{\nu}k_{\mu}-(p_{2})_{\mu}(p_{1})_{\nu}\Big]\frac{({\epsilon_{1}}^{\mu})^{\ast}({\epsilon_{2}}^{\nu})^{\ast}}{D(q_{2}^{2}-M^{2}_{w})}.
\end{align} 

Therefore, it can be easily seen that the terms $M_{\mu\nu}^{(a)}$, $M_{\mu\nu}^{(b)}$, and $M_{\mu\nu}^{(c)}$ are given by
\begin{align}
M_{1}^{(-2;div)}=M_{\mu\nu}^{(a)}({\epsilon_{1}}^{\mu})^{\ast}({\epsilon_{2}}^{\nu})^{\ast},\\
M_{t}^{(0,div)}=M_{\mu\nu}^{(b)}({\epsilon_{1}}^{\mu})^{\ast}({\epsilon_{2}}^{\nu})^{\ast},\\
M_{t}^{(0,fin)}=M_{\mu\nu}^{(c)}({\epsilon_{1}}^{\mu})^{\ast}({\epsilon_{2}}^{\nu})^{\ast}.
\end{align}

\end{document}